\documentclass[prx,twocolumn,superscriptaddress,longbibliography]{revtex4-2}
\usepackage{bm}
\usepackage{amsmath, amsfonts}
\usepackage{amssymb}
\usepackage{times}
\usepackage{graphicx}
\usepackage{svg}
\usepackage{braket}
\usepackage{xcolor}
\usepackage{mathtools}
\usepackage{float}
\usepackage[colorlinks=true,allcolors=blue]{hyperref}
\usepackage[capitalise,nameinlink]{cleveref}
\usepackage{tikz}
\crefname{section}{Sec.}{Figs.}
\graphicspath{{figures/}}
\begin{document}

\title{WSi weak link element with a non-sinusoidal current-phase relation}

\graphicspath{ {./figures/} }

\author{Sarah Garcia Jones}
\thanks{These authors contributed equally to this work.}
\affiliation{Department of Physics, University of Colorado Boulder, Boulder, CO 80309, USA}
\affiliation{Department of Electrical, Computer \& Energy Engineering, University of Colorado Boulder, Boulder, CO 80309, USA}

\author{Trevyn F. Q. Larson}
\thanks{These authors contributed equally to this work.}
\affiliation{Department of Physics, University of Colorado Boulder, Boulder, CO 80309, USA}
\affiliation{National Institute of Standards and Technology, Boulder, CO 80305, USA}

\author{Sai Pavan Chitta}
\thanks{These authors contributed equally to this work.}
\address{Department of Physics and Astronomy, Northwestern University, Evanston, IL 60208, USA}

\author{Heli Vora}
\affiliation{National Institute of Standards and Technology, Boulder, CO 80305, USA}

\author{Varun Verma}
\affiliation{National Institute of Standards and Technology, Boulder, CO 80305, USA}

\author{\\Sae Woo Nam}\thanks{Deceased}
\affiliation{National Institute of Standards and Technology, Boulder, CO 80305, USA}

\author{José Aumentado}
\affiliation{National Institute of Standards and Technology, Boulder, CO 80305, USA}

\author{Jens Koch}
\address{Department of Physics and Astronomy, Northwestern University, Evanston, IL 60208, USA}

\author{Raymond W. Simmonds}
\email{raymond.simmonds@nist.gov}
\affiliation{National Institute of Standards and Technology, Boulder, CO 80305, USA}

\author{Andr\'as Gyenis}
\email{andras.gyenis@colorado.edu}
\affiliation{Department of Electrical, Computer \& Energy Engineering, University of Colorado Boulder, Boulder, CO 80309, USA}
\affiliation{Department of Physics, University of Colorado Boulder, Boulder, CO 80309, USA}

\begin{abstract}
\label{sec:abstract}
Nonlinearity is an essential ingredient for encoding quantum states with non-uniform energy spacing, implementing coherent quantum gates, reading out qubits, amplifying, and mixing electromagnetic signals. In this work, we demonstrate the nonlinear behavior of a constriction fabricated from an amorphous, high-kinetic inductance material, tungsten silicide, embedded in a three-dimensional RF-SQUID. We find that the results are consistent with the weak link behaving as a Josephson junction with a sawtooth-like current-phase relation or a quantum phase slip element. Finally, we measure relaxation times of the metastable, persistent-current states trapped in the local minima of the potential.
\end{abstract}

\maketitle 

\section{Introduction}
\label{sec:introduction}

For the practical implementation of quantum computers, it is essential to isolate a two-dimensional subspace within a multilevel quantum system.  This can be done, for example, by eliminating a degree of freedom of the physical system, such as the orbital component of a wavefunction, or by truncating the number of energy eigenstates based on their non-equidistant spacing. Nonlinearity in quantum systems is thus a key resource for implementing qubits in physical systems.  While atoms and ions are naturally nonlinear, quantum circuits built from superconducting capacitors and inductors require an additional nonlinear circuit element to encode qubits. Such nonlinearity can arise from altering the superconducting properties at a certain position in the circuit. For example, inserting non-superconducting barriers between superconductors can generate nonlinearity due to the Josephson effect~\cite{josephson_possible_1962,Kockum2019}. The choices for barrier materials can be, for instance, insulators~\cite{Martinis2020}, semiconductors~\cite{de_lange_realization_2015, larsen_semiconductor-nanowire-based_2015, pita-vidal_gate-tunable_2020, ramon_2020}, ferromagnets~\cite{PhysRevLett.104.137002,PhysRevB.107.L081301}, two-dimensional electron gases~\cite{casparis_superconducting_2018}, van der Waals materials~\cite{wang_coherent_2019, liu_2d_2019}, or topological compounds~\cite{Ren2019}.  Among the numerous options, Josephson junctions made from oxidized aluminum are the most popular option nowadays in superconducting qubits~\cite{annurev}. However, these junctions can host unwanted two-level fluctuators in the tunnel barrier~\cite{PhysRevLett.95.210503}, cause unpredictable circuit parameters~\cite{hertzberg_laser-annealing_2021}, have parasitic capacitances that burden the realization of protected devices~\cite{gyenis_2021}, and {restrict the operation temperature of superconducting qubits~\cite{PRXQuantum.5.030347}.}

An alternative route to nonlinearity is to avoid introducing different barrier materials and rely on superconducting weak links~\cite{likharev_superconducting_1979}. These weak links are nanoscale constrictions with dimensions comparable to the coherence length and have different superconducting properties than the rest of the circuit. In weak links, the supercurrent flows along a conducting path, where the critical current differs from that of the electrodes connected to it. Such non-tunnel-type structures can exhibit weak link Josephson junction effects, where the current phase relation (CPR) might be strongly non-sinusoidal~\cite{golubov_current-phase_2004}. Recently, deviation from a  {purely} sinusoidal CPR due to higher harmonics has been the focus of several studies concerning the channel transmission physics in traditional junctions~\cite{Willsch2024, kim_emergent_2025}. Understanding and taking advantage of these higher harmonics is of interest since it can support the realization of protected qubits~\cite{PRXQuantum.3.030303,bozkurt2023, PhysRevLett.125.056801, feldsteinbofill2026controlledparitycooperpair} and reduction of offset-charge-sensitivity~\cite{PhysRevLett.124.246803,PhysRevLett.124.246802}. In weak link nanowires, quantum phase slip events can also arise when the phase fluctuations are strong~\cite{astafiev_coherent_2012}. In that case, the weak link is modeled by a circuit element that is dual to the Josephson junction, describing the coherent tunneling of fluxons~\cite{Mooij2006}. Weak links have been realized using thin-film evaporation and etching~\cite{astafiev_coherent_2012}, focused-ion-beam milling~\cite{potter_controllable_2023, PhysRevApplied.11.014006}, atomic layer etching~\cite{bottcher2025transmonqubitrealizedexploiting}, atomic contacts~\cite{janvier_coherent_2015}, and fabricated in various materials, for instance, in aluminum~\cite{vijay_optimizing_2009,vijay_approaching_2010,levenson-falk_nonlinear_2011}, granular aluminum~\cite{winkel_implementation_2020, schon_rabi_2020, rieger_granular_2023}, InO$_x$ \cite{astafiev_coherent_2012}, TiN~\cite{de_graaf_charge_2019, purmessur2025}, and NbN~\cite{shaikhaidarov_quantized_2022, peltonen_coherent_2013,de_graaf_charge_2018, bottcher2025transmonqubitrealizedexploiting}.

In this work, we present the behavior of a radio-frequency superconducting quantum interference device (RF-SQUID) containing a tungsten-silicide (WSi) nanowire as the nonlinear weak link element.  {WSi is an amorphous disordered superconductor that has a critical temperature around $4\,$K, and a coherence length of approximately $7\,$nm, which depends on the film thickness and stoichiometry~\cite{Kondo1992}. This material} has been the primary choice for superconducting nanowire single-photon detectors because it has extreme structural homogeneity, which allows the fabrication of uniform nanowire detectors with a large number of pixels~\cite{marsili_detecting_2013,oripov_superconducting_2023}. Furthermore, WSi has also been implemented as a linear inductor in fluxonium qubits~\cite{Larson2026}.  Here, we use WSi as a nonlinear nanowire inside a superconducting loop shunted by a large capacitor. Using microwave spectroscopy, we observe a flux-dependent resonance frequency, which can be captured by modeling the weak link either by a Josephson junction with a strongly non-sinusoidal CPR or a quantum phase slip element.  {We theoretically model the device with the Josephson junction based on the traditional approach to circuit quantization~\cite{https://doi.org/10.1002/cta.2359}. To obtain the Hamiltonian of the system with the quantum phase slip element, however, we harness a recently developed technique based on the symplectic geometry of the circuit~\cite{PRXQuantum.5.020309}.} Using single-shot readout, we measure hour-long relaxation times of the excited fluxon states of the RF-SQUID due to the reduced transition matrix elements between the eigenstates. This work demonstrates the feasibility of WSi as a nonlinear material in superconducting circuits, opening up possibilities to realize weak-link quantum devices.

This paper is organized as follows. In Section~\ref{sec:weaklink}, we review two circuit representations of a weak link: a Josephson junction and a quantum phase slip element. In Section~\ref{sec:measurement}, we describe the physical WSi weak-link device and present its experimental resonance condition under changing magnetic flux. Next, we explain the theoretical description of the measurements in Section~\ref{sec:theory_model}. Finally, we discuss the time-domain measurements of the metastable states in the RF-SQUID in Section~\ref{sec:time_domain}.

\section{The Weak link circuit element}
\label{sec:weaklink}

\begin{figure}[b]
    \centering
    \includegraphics[width = 8.6cm]{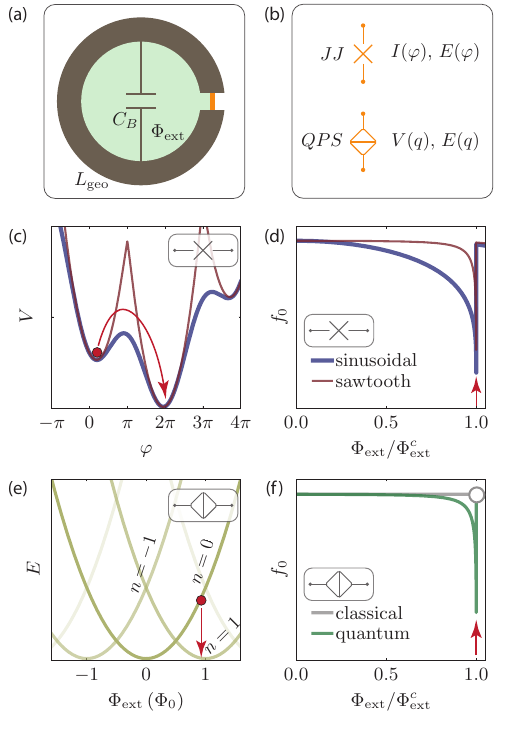}
    \caption{\label{fig:intro} (a) Schematics of a flux-tunable LC-circuit with a weak link (orange). The geometric inductance of the loop due to the superconducting wire (brown) is $L_\mathrm{geo}$, and the shunting capacitance is $C_B$. The green region indicates the external flux $\Phi_\mathrm{ext}$ piercing the loop. (b) Two different circuit elements modeling the weak link: a JJ with current-phase relation $I(\varphi)$ and energy-phase relation $E(\varphi)$, or a QPS with voltage-charge relation $V(q)$ and energy-charge relation $E(q)$. (c) In the JJ-description, the potential from a sinusoidal CPR has a corrugated shape (dark blue), while a sawtooth CPR results in a similar shape but with much higher well barriers (dark red). Red arrows indicate the process when the system leaves a local minimum in the potential. (d) Model resonance curves as a function of flux for the sinusoidal and sawtooth-like CPR. (e) The energy of the system as a function of flux for different numbers of enclosed flux quanta $n$ when the weak link is modeled as a QPS. The system can decay to a lower state through a phase slip process (red arrow). (f) Model resonance curves as a function of flux when the phase slips are treated as classical and quantum processes. The white dot indicates that the classical resonance frequency is undetermined at the critical flux value.}
\end{figure}

To capture the concept behind the experiment in this work, we first consider a simpler circuit: a flux-tunable LC-resonator, which is formed by a superconducting loop with geometric inductance $L_\mathrm{geo}$ and shunt capacitance $C_B$ [Fig.~\ref{fig:intro}(a)]. The loop contains a narrow weak link (a nanowire) and is placed in an external magnetic field $\Phi_\mathrm{ext}$. Our goal is to understand the behavior of the weak link and model it with a lumped circuit element. In general, we can consider two circuit representations of the weak link: either a Josephson junction (JJ) or a quantum phase slip (QPS) junction [Fig.~\ref{fig:intro}(b)].

First, we can think of the constriction as a barrier for Cooper pairs and describe it with weak-link Josephson physics~\cite{likharev_superconducting_1979}. In this case, the nanowire adds a tunable nonlinear inductance $L_\mathrm{nw}$ to the loop inductance, leading to an inductive energy $V(\varphi)$ with a corrugated shape as a function of phase $\varphi$ across the weak link [Fig.~\ref{fig:intro}(c)]. In a semiclassical picture, where the system is approximated by a fictitious particle moving in the potential landscape, the local curvature of the potential determines the effective inductance of the circuit and its resonance frequency. In this approximation, the mass of the particle is set by the capacitance, while the linear and Josephson inductance determine the potential landscape. By changing the external flux and tuning the shape of the potential, we can gain information about the CPR of the nanowire. For example, the difference between sinusoidal and sawtooth-like CPR results in resonance frequencies with vastly different shapes [Fig.~\ref{fig:intro}(c) and (d)]. Furthermore, if the ratio of the inductances $\beta=L_\mathrm{geo}/L_\mathrm{nw}>1$, the system has multiple local minima where the particle can be localized. When the system leaves such a metastable state around a critical flux $\Phi_\mathrm{ext}^c$, the resonance frequency exhibits a sudden change and hysteretic behavior [red arrows in Fig.~\ref{fig:intro}(c) and (d)]. 

Second, we can consider an alternative picture to explain the behavior of the device due to phase slips. Initially, when we increase the external flux in the loop from zero flux, a circulating current $I$ is induced that ensures that the total magnetic flux inside the loop $\Phi_\mathrm{inside}$ remains zero to satisfy flux quantization. This leads to an increase in the inductive energy $E_\mathrm{ind}$, because $E_\mathrm{ind}= LI^2/2=\Phi_\mathrm{ext}^2/2L$, where $L$ is the total inductance of the loop. In general, when there are $n$ flux quanta enclosed in the loop, $\Phi_\mathrm{inside}=n\Phi_0$, the total inductive energy is $E_\mathrm{ind}= (\Phi_\mathrm{ext}-n\Phi_0)^2/2L$ [Fig.~\ref{fig:intro}(e)]. Notice that this is the electromagnetic dual of the charging energy of Cooper pairs on a superconducting island. Thus, upon raising the external flux, it is energetically favorable for the system to adjust the enclosed flux in the loop to a multiple of flux quanta that is closer to the external flux [red arrow in Fig.~\ref{fig:intro}(e)]. This process occurs at  {a critical flux value $\Phi_\mathrm{ext}^c$} via a flux quantum entering the loop, accompanied by a $2\pi$ phase shift of the superconducting order parameter and the temporary closing of the superconducting gap~\cite{Mooij2006}. Hence, this event is commonly referred to as a phase slip event. The  {physical} position where a phase slip is most likely to occur is at the  {``weak link'' of the circuit}, because the fluctuation of the superconducting order parameter is enhanced in the constriction. Additionally, in weak links fabricated from high-kinetic inductance materials, the phase stiffness is reduced, which further enhances the possibility of phase slips~\cite{Charpentier2025}.  {This is similar to the case when a Josephson junction is placed in a high impedance environment, where the low-energy spectrum of the circuit can be approximated by a phase slip model~\cite{PhysRevX.14.041014}.} In a semiclassical treatment  {of phase slips}, such an event only alters the resonance frequency of the device when the phase slip occurs; otherwise, the resonance frequency remains constant because the inductance of the circuit is unchanged [Fig.~\ref{fig:intro}(f)]. In a quantum modeling of the phase slip event, however, it is possible that in the vicinity of the phase slip, the resonance frequency decreases due to the tails of the quantum wavefunctions, as we discuss in Sec.~\ref{sec:theory_model}.

We note that the two models predict a similar flux-dependence for the resonance curve as long as the Josephson junction has a sawtooth-like CPR (or a quadratic energy phase relation). This is because, as the flux evolves, in both cases, the system experiences the same effective constant curvature in the potential, leading to a relatively flat response of the resonance. In this work, we experimentally show this behavior in an RF-SQUID circuit.

\section{Measurement of the resonance condition}
\label{sec:measurement}

\begin{figure}[t]
\centering
\includegraphics[width = \linewidth]{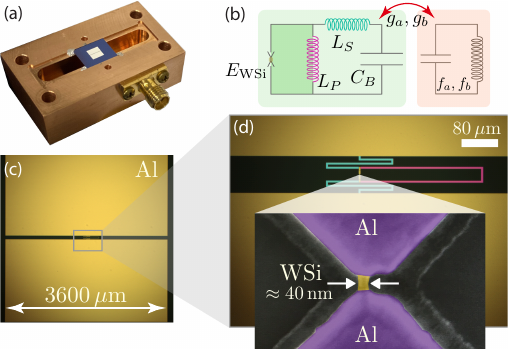}
\caption{\label{fig:device} (a) RF-SQUID circuit coupled to a three-dimensional copper cavity. The RF-SQUID is patterned on a silicon chip suspended in a slot that is machined into the copper cavity and enclosed by an identical copper piece (not pictured). (b) Lumped element circuit diagram. The red arrow refers to the coupling $g_a$ and $g_b$ between the cavity (orange) and the RF-SQUID (light green). The RF-SQUID device is formed by a loop containing the weak link with energy $E_\mathrm{WSi}$ and a parallel inductor $L_p$ (pink). In addition, the device has a series inductor $L_s$ (cyan), and a large capacitor $C_B$. (c) Optical microscope image of the RF-SQUID device.  The capacitor pads (yellow rectangles) are much larger than the meandering inductor between them (each pad measures $1.8\times3.6$\,mm). (d) False colored optical microscope image of the inductor and the loop between the capacitor pads. Parallel inductance $L_p$ is shown in pink while series inductance $L_s$ is shown in cyan. The zoomed-in inset is a false colored scanning electron microscope image of the WSi nanowire (yellow), defined between triangular aluminum contacts (purple).  {The shadow around the triangular aluminum contacts is the residual of the electron beam resist from the lithographic write.}}
\end{figure}

Our device contains an on-chip RF-SQUID circuit that is strongly coupled to a three-dimensional copper cavity [Fig.~\ref{fig:device}(a) and (b)]. The circuit is formed by a large shunting capacitor, an aluminum loop containing the WSi nanowire, and series inductors [Fig.~\ref{fig:device}(c) and (d)]. We fabricated the WSi nanowire with approximately $40\times 40$\,nm lateral dimensions by depositing and etching a 3\,nm thick WSi film with a kinetic sheet inductance of $L_K\approx 300\,\mathrm{pH}/\square$ (see details in App.~\ref{app:fabrication}). We used a lift-off processes to fabricate the large aluminum capacitor pads and connecting wires. Based on  {low-frequency} electromagnetic simulations, the shunting capacitance across the loop is $C_B=646$\,fF. The inductance of the wire parallel with the weak link has a value of $L_p= 362$\,pH, while the inductance in series with the nanowire has a value of $L_s=288$\,pH. The large capacitor pads provide a strong coupling to the TE$_{101}$ and TE$_{301}$ modes of the cavity, which hybridize with the on-chip mode of the RF-SQUID. Details on the renormalized cavity frequencies, $f_a$ and $f_b$, as well as the coupling between the modes and the
RF-SQUID, $g_a$ and $g_b$, are available in App~\ref{app:finite-element}. In this work, we measure and model the hybridized resonance frequency of the coupled system [Fig.~\ref{fig:device}(b)]. An attached external coil provides the magnetic flux for the device. 

\begin{figure}[t]
    \centering
    \includegraphics[width = 8.6cm]{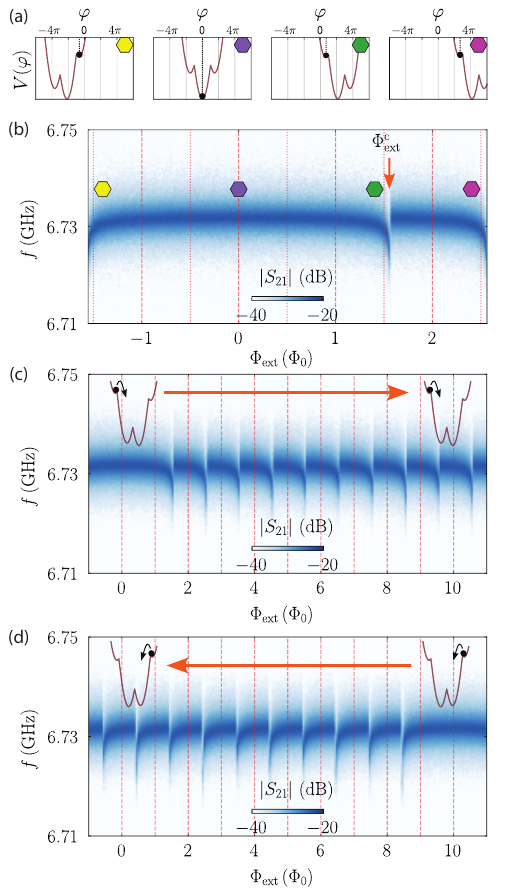}
    \caption{\label{fig:resonance_measurement}(a) A series of toy-model potential landscapes as the external flux evolves.  The state of the system is indicated by a phase particle (black dot) in a local minimum in each panel. (b) Transmission data showing the resonance frequency as a function of applied magnetic field and frequency $f$.  The colored hexagons at various flux values refer to the potentials in (a). In this data set, flux is increased from around $-1.5$ to $2.5\,\Phi_0$, with a phase jump occurring at approximately $1.56\,\Phi_0$, labeled as $\Phi^\mathrm{c}_{\mathrm{ext}}$. The curvature of the resonance frequency response is relatively flat rather than sinusoidal, suggesting a more sawtooth-like current phase relation [see Fig.~\ref{fig:intro}(d)].  (c), (d) Resonance frequency data showing the hysteretic response of the RF-SQUID to increasing and decreasing external magnetic flux over a larger interval.} 
\end{figure}

We begin by measuring the magnitude of the transmission $|S_{21}|$ across the cavity while ramping the applied flux from around $-1.5\,\Phi_0$. As Fig.~\ref{fig:resonance_measurement} (b) shows, after an initial upward curve,  the resonant frequency plateaus at a stable value of around 6.73\,GHz over multiple flux quanta. When we approach a critical flux value $\Phi_\mathrm{ext}^c\approx 1.56\,\Phi_0$, the resonance curve decreases by several MHz before sharply returning to its original value. After this initial jump, continuing to change the magnetic field in the same direction results in multiple jumps with periodicity of one flux quantum [Fig.~\ref{fig:resonance_measurement}(c)]. When we reverse the direction of the flux sweep, the jumps show a mirrored behavior, signaling a hysteresis in the frequency-flux scan [Fig.~\ref{fig:resonance_measurement}(d)]. 

We first explain this behavior based on a semiclassical toy model, followed by a more detailed description in Section~\ref{sec:theory_model}. When we treat the weak link as a Josephson junction in the RF-SQUID, the potential has a corrugated structure due to the presence of an inductor and the junction. The shape of the potential can be tuned by the magnetic field, as displayed in Fig.~\ref{fig:resonance_measurement}(a). Given the large shunting capacitance and the large junction energy, we can describe the state of the system as a phase particle, represented by a black dot in Fig.~\ref{fig:resonance_measurement}(a). In this picture, the resonance frequency corresponds to the oscillation frequency of the particle, which depends on the local curvature of the well and the mass of the particle (capacitance of the device). Note that the sawtooth CPR gives rise to a potential consisting of an infinite series of parabola-like valleys superimposed on a larger parabola.

We initialize the system with the particle resting at the bottom of the global minimum of the potential. From here, we can ramp the field in either direction without significantly changing the resonance frequency as long as the phase particle remains trapped in its original well and the curvature  {of the potential} is mostly unchanged. When we ramp the field beyond a half flux quantum, the central well is lifted above its neighbor, but owing to the large potential barrier, the phase particle remains trapped in the same minimum. Upon further increase of the field, the potential barrier between the wells is reduced enough that the particle can escape its original well and jump into the adjacent well. This event manifests itself as a sudden change in the resonance frequency. We note that just before this jump occurs, the curvature of the potential changes, leading to a reduced resonance frequency. When the flux ramping direction is reversed, the well containing the phase particle must decrease to a minimum before again elevating above its other adjacent wells, hence the longer flux range before the periodic motion begins in the opposite direction [Fig.~\ref{fig:resonance_measurement}(d)].

\section{Theoretical modeling of the circuit}
\label{sec:theory_model}

\begin{figure}[t!]
    \centering
    \includegraphics[width = 8.6cm]{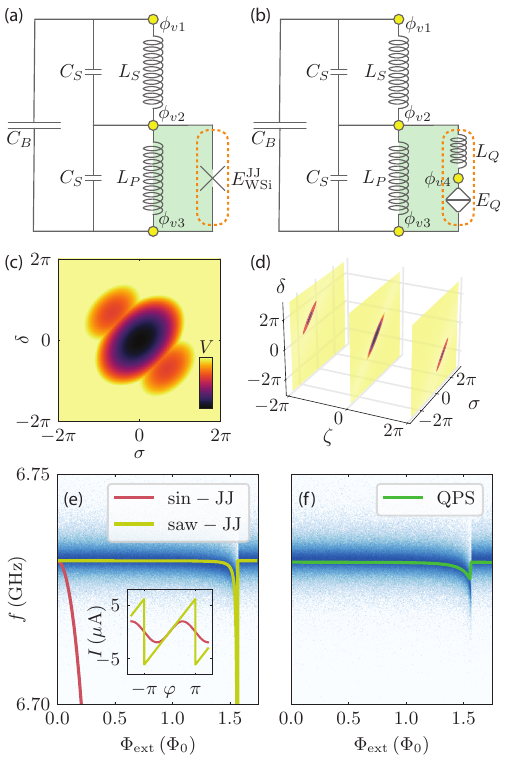}
    \caption{\label{fig:theory} The lumped-element circuit model of the device, when the nanowire (dotted orange rectangle) is treated (a) as a JJ or (b) as a QPS. (c), (d) The potential energy of the circuits in the two cases (a) and (b), respectively. Because the circuit has three degrees of freedom in the QPS case, the configuration space is three-dimensional. The colorbar ranges from 0 to $10^4$\,GHz in (c) and from 0 to $2\cdot10^4$\,GHz in (d). (e), (f) The calculated resonance curves of the two circuits superimposed on the experimental data. The inset of (e) shows the corresponding CPRs of each JJ model. In (e),  {the parameters} are $E_{LP}/h= 452\,$GHz, $E_{LS}/h= 568\,$GHz, $E_{C\sigma}/h=7.4\,$MHz, $E_{C\delta}/h=0.97\,$GHz, where the parasitic capacitance is estimated to be $C_S=10$\,fF. By varying the coupling constants to the modes of the cavity, the junction energy and shape, we find  {$E^0_\mathrm{WSi}/h=973\,$GHz}. Furthermore, $\chi=0$ for the sinusoidal case,  {$\chi=1-5\cdot10^{-5}$} for the sawtooth case, and  {$g_a/2\pi=1.73\,$GHz and $g_b/2\pi= 2.12\,$GHz}. For the QPS model, the parameters are the following: $E_{LP}/h= 305\,$GHz, $E_{LS}/h= 840\,$GHz, $E_{C\sigma}/h=5.2\,$MHz, $E_{C\delta}/h=0.97\,$GHz, $E_Q/h=60\,$GHz, $L_Q=8\,$pH, and $g_a/2\pi=2.19\,$GHz and $g_b/2\pi= 2.67\,$GHz.   {We used the results of finite-element simulations as the initial guess for the parameters (App.~\ref{app:finite-element}). Because the device has large elements on the scale of the wavelength of the resonance, the result of a DC simulation can be considered only as an estimate for the parameters of the lumped circuit elements.} In both models, these parameters are not unique to match the data.}
\end{figure}

We use two conceptually distinct lumped-element circuit models to describe the measured resonance frequency [Fig.~\ref{fig:theory}(a) and (b)]. The models follow the different interpretations of how a constriction in a superconducting wire can be modeled: as a Josephson junction or a quantum phase slip element. We note that while both models agree with the experimental data in the realized parameter regime, the measured resonance alone is insufficient to distinguish between the two possibilities because the low-energy spectra of the two circuit models are similar  {(App.~\ref{app:full_spectrum})}.

In the first case [Fig.~\ref{fig:theory}(a)], we model the constriction as a weak-link Josephson junction, where the element has an energy-phase constitutive relation $E_\mathrm{WSi}^\mathrm{JJ}(\varphi)$, which is a periodic function of the superconducting phase $\varphi$ across the junction. When the transmission probability of the Cooper pairs across the weak link is very low, as in aluminum-oxide tunnel barriers, this periodic function is well-described with a sinusoidal function~\cite{Willsch2024}. When the Cooper pairs' probability of passing through the barrier is larger, higher harmonics appear in the energy-phase function, resulting in more skewed CPRs~\cite{vijay_optimizing_2009,PhysRevLett.125.056801}. 

Using standard circuit quantization techniques~\cite{https://doi.org/10.1002/cta.2359}, we arrive at the Hamiltonian of the circuit
\begin{equation}\label{eq:Hamiltonian_JJ}
\begin{split}
\hat{H}^\mathrm{JJ} &=  4E_{C\sigma}\hat{n}_\sigma^2 + 4E_{C\delta}\hat{n}_\delta^2 + \\
&\frac{1}{2}E_{LS}\cdot\left(\hat{\sigma}+\hat{\delta}\right)^2+\frac{1}{2}E_{LP}\cdot\left(\hat{\sigma}-\hat{\delta}-2\pi\frac{\Phi_\mathrm{ext}}{\Phi_0}\right)^2 + \\
& E_\mathrm{WSi}^\mathrm{JJ}(\hat{\delta}-\hat{\sigma}),
\end{split}
\end{equation}
where $\hat{\sigma}$ and $\hat{\delta}$ refer to the common (heavy) and differential (light) mode operators, and $\hat{\sigma}=(\hat{\phi}_3-\hat{\phi}_1)/2$ and $\hat{\delta}= (2\hat{\phi}_2 - \hat{\phi}_1 - \hat{\phi}_3)/2$, where $\hat{\phi}_i$ are the node phase operators in the circuit ($i=1,2,3$). The corresponding conjugate Cooper pair number operators are $\hat{n}_\sigma$ and $\hat{n}_\delta$. The inductive energies are $E_{LS}= \Phi_0^2/4\pi^2L_S$ and $E_{LP}= \Phi_0^2/4\pi^2L_P$, while the charging energies are $E_{C\delta}=e^2/4C_S \gg E_{C\sigma}=e^2/4(C_S+2C_B)$, where $C_S$ is the small parasitic capacitance between the nodes. To describe the weak link, we consider two different CPRs: a sinusoidal and a skewed one, both of which can be mathematically written as an energy-phase relation of the form
\begin{equation}
        E_\mathrm{WSi}^\mathrm{JJ}(\varphi)  = -E_\mathrm{WSi}^0(1 +\chi)\sum_{k=1}^\infty(-\chi)^{k-1}\frac{\cos(k\varphi)}{k^2}.
\end{equation}
Here, $E_\mathrm{WSi}^0$ is the characteristic energy scale of the junction, and we introduce $\chi$ as the degree of skewness, with $\chi=0$ leading to the purely sinusoidal, while $\chi=1$ to the sawtooth CPR. Figure~\ref{fig:theory}(c) shows an example of the two-dimensional potential energy of the circuit.

Given the large shunt capacitance and large junction energy, we model the behavior of the resonance frequency using a semiclassical approximation. In this case, we calculate the curvature of the potential minima in phase space as a function of the external flux to obtain an effective inductance and resonance frequency of the circuit.  When modeling the weak link as a Josephson junction with a purely sinusoidal CPR, the theoretical curve fails to capture the experimental data [red line in Fig.~\ref{fig:theory}(e)]. In contrast, when assuming a sawtooth-like CPR for the weak link, the calculated results describe the measurements with significantly larger accuracy [yellow line in Fig.~\ref{fig:theory}(e)]. In this model, the downturn of the resonance near the jump comes from the curving of the sawtooth-like CPR close to $\pi$ phase.  {We note that the theoretical curve predicts a critical flux value where the curvature of the local minima vanishes. Thus, it overestimates the critical flux value because the jump between the valleys can occur at a lower flux due to either quantum tunneling or thermal excitation.} 

In the second case, we model the weak link as a quantum phase slip element [Fig.~\ref{fig:theory}(b)]. When phase slip events are coherent, the nanowire can be described by a quantum phase slip junction, which has an energy-charge constitutive relation~\cite{Mooij2006}. In the simplest case, this can be written as $E_\mathrm{WSi}^\mathrm{QPS}(\hat{n}) = -E_Q\cos(2\pi \hat{n})$, where $E_Q$ is the amplitude of the quantum phase slip events, and $\hat{n}$ is the Cooper pair number operator. We show the corresponding circuit diagram for the device in Fig.~\ref{fig:theory}(b), where we included the parasitic inductive component $L_Q$ in series with the QPS element. This is dual to the case of a Josephson junction, where a parasitic capacitor in parallel to the junction needs to be included to avoid singularities in the circuit~\cite{PhysRevX.13.021017}. Analyzing such a circuit is more unconventional, and arriving at the corresponding Hamiltonian requires the use of loop charges~\cite{PhysRevB.94.094505} or the description of the symplectic structure associated with the capacitive subgraph of the circuit~\cite{PRXQuantum.5.020309}. We describe the details of the calculation in App.~\ref{app:circuit_quantization}, which leads to the Hamiltonian
\begin{equation}\label{eq:qps_hamiltonian}
\begin{split}
\hat{H}^\mathrm{QPS} &= 4E_{C\sigma}\hat{n}_\sigma^2 + 4E_{C\delta}\hat{n}_\delta^2 - E_\mathrm{WSi}^\mathrm{QPS}\cos\left(2\pi\hat{n}_\zeta\right) +  \\
&\frac{1}{2}E_{LS}\cdot\left(\hat{\sigma}+\hat{\delta}\right)^2+\frac{1}{2}E_{LP}\cdot\left(\hat{\sigma}-\hat{\delta}-2\pi\frac{\Phi_\mathrm{ext}}{\Phi_0}\right)^2 + \\
& \frac{1}{2}E_{LQ}\cdot\left(\hat{\sigma}-\hat{\delta}-\hat{\zeta}\right)^2.
\end{split}
\end{equation}
This circuit has three modes due to the additional node, and besides the common and differential modes defined for the Josephson-junction-based circuit, here, the $\hat{\zeta}=\hat{\phi}_4-\hat{\phi}_3$ mode operator corresponds to the number of $2\pi$-phase slips across the QPS, and it is defined on the integer multiples of $2\pi$. The cosine of the conjugate charge operator is $\cos\left(2\pi\hat{n}_\zeta\right)$. In addition to the quantities defined in Eq.~\ref{eq:Hamiltonian_JJ},  here $E_{LQ}= \Phi_0^2/4\pi^2L_Q$. Figure \ref{fig:theory}(d) shows the potential energy of the circuit in the three-dimensional configuration space. We use perturbation theory to obtain the resonance frequency of this circuit (App.~\ref{app:circuit_quantization}), which leads to agreement with the experimental data [Fig.~\ref{fig:theory}(f)].  {Critically, the downturn of the resonance frequency in this case can be understood based on a quantum mechanical coupling between neighboring wells, unlike in the first, semiclassical model, where the decrease in the frequency arises from the change of the shape of the current-phase-relation. In this theoretical description, we calculate the eigenstates of the Hamiltonian without taking into account the decay between valleys. The exact location of the critical flux is probabilistic and depends on the loss model used and the rate of flux ramp. Here, we approximate its location based on perturbation theory: we choose a critical flux value corresponding to a coupling rate between states in neighboring valleys that is on par with the flux ramp rate (see details in the App.~\ref{app:circuit_quantization}).  While this approximation certainly does not provide an exact description of the critical flux, this Hermitian model captures the curve up to the close vicinity of the flux jump.} Finally, we note that, due to the number of free parameters in the models, multiple combinations of values can lead to indistinguishable resonance curves for both models. Thus, these measurements alone are not suitable to determine the precise values of the parameters in the Hamiltonian.

\section{Time-domain measurements of the metastable states}
\label{sec:time_domain}

\begin{figure}[b]
    \centering
    \includegraphics[width = 8.6cm]{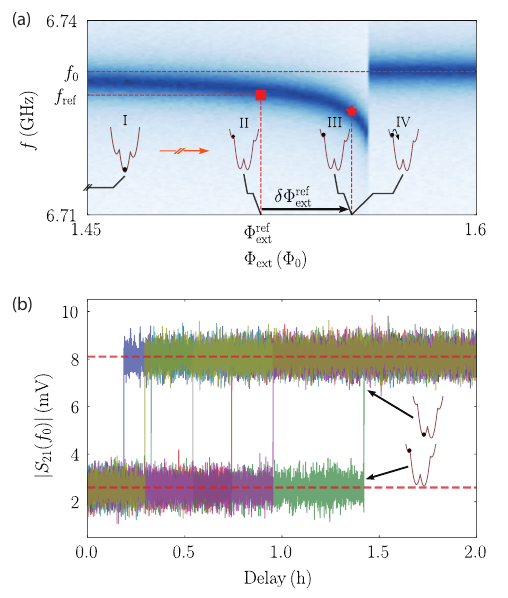}
    \caption{\label{fig:time_domain}(a) Example of the flux-dependent resonance curve featuring a single switching event. The steps of our time-domain measurement procedure are overlaid at each flux point of interest. The steps are as follows: (I) Initialization of the system in a global minimum of the potential at a flux value $\Phi_\mathrm{ext}^\mathrm{init}$ (not shown on this plot). (II) Coarse ramping of the flux to $\Phi_\mathrm{ext}^\mathrm{ref}$, identified in the measurement when the resonance peaks at $f_\mathrm{ref}$. This setpoint approaches the critical flux value but is far enough that the likelihood of a jump is negligible. (III) Fine ramp of the flux by $\delta\Phi_\mathrm{ext}^\mathrm{ref}$, where the time measurement is carried out. (IV) Single-shot time-domain measurement at $\Phi_\mathrm{ext}^\mathrm{ref} + \delta\Phi_\mathrm{ext}^\mathrm{ref}$ to detect a switching event. When this occurs, the resonance frequency returns to its original higher value $f_0$. (b)  {Examples of raw single-shot data obtained using the same fine detuning $\delta\Phi_\mathrm{ext}^\mathrm{ref}$ that show switching events with lifetimes over an hour.}}
\end{figure}

Now, we turn to the time-domain measurements to determine the lifetime of the prepared persistent-current states. We use the weak-link junction interpretation of the nanowire to describe the time-domain measurements, though a similar description may be applied using the QPS interpretation as well. As described in Section \ref{sec:theory_model}, the device has a large characteristic junction energy $E^0_\mathrm{WSi}\approx1055\,\mathrm{GHz}$ and a small charging energy $E_{C\sigma}\approx7\,\mathrm{MHz}$.  This combination of energy scales strongly localizes the states in different valleys, resulting in reduced matrix elements and long relaxation rates of the flux states~\cite{PRXQuantum.2.030101,lin2018,earnest2018,Hassani2023}. 

\begin{figure}[b]
    \centering
    \includegraphics[width = 8.6cm]{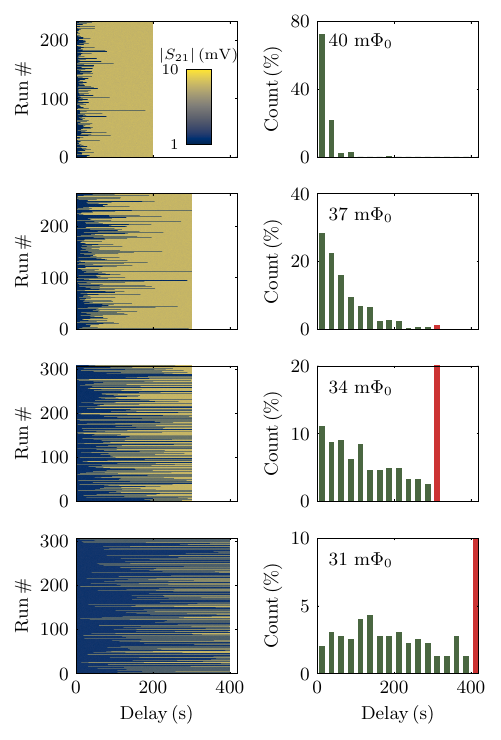}
    \caption{\label{fig:histograms} Histograms showing measured lifetimes at decreasing $\delta\Phi_\mathrm{ext}^\mathrm{ref}$ values (increasing potential barriers). Plots on the left show individual single-shot experiments, similar to Fig.~\ref{fig:time_domain}(b), with identical state preparations in each panel where $|S_{21}(f_{0})|$ is continuously measured. Here, blue (low $|S_{21}(f_0)|$ signal) corresponds to the excited state, and gold (high $|S_{21}(f_0)|$ signal) corresponds to the state after decay. Plots on the right of each panel show corresponding histograms where each bin represents a different lifetime. As $\delta\Phi_\mathrm{ext}^\mathrm{ref}$ decreases, the barrier between neighboring wells increases, leading to longer lifetimes. The red bars represent counts where the state did not decay in the measurement window, meaning that the lifetimes were longer than the length of the measurement.  {(The heights of the red bars are 28\% and 59\% in the bottom two histograms, from top to bottom.)}} 
\end{figure}

To demonstrate the presence of long-lived excited states, we carry out time-domain measurements to monitor the decay of localized metastable states through different barrier heights from one flux state to another. The measurement protocol consists of four steps, illustrated in the inset of Fig.~\ref{fig:time_domain}(a). First, we initialize the system in its ground state, where the particle rests at the bottom of the lowest valley. Second, we ramp the flux such that this valley rises above its neighboring valley. We tune the flux slowly enough to avoid exciting the state out of the valley, but fast enough that the state has a low probability of escaping during the ramp. To consistently set an initial flux value for every run, we continuously monitor the amplitude of the transmission value $|S_{21}|$ at a specific reference frequency,  $f_\mathrm{ref} = 6.7285$\,GHz, which lies just before the frequency plummets. This corresponds to a flux of $\Phi_\mathrm{ext}^\mathrm{ref} = 1.516\,\Phi_0$ (red square in Fig.~\ref{fig:time_domain}).  From this reference point, step three is to detune the flux once more with a smaller value of $\delta\Phi_\mathrm{ext}^\mathrm{ref}$ and carry out the time-domain measurement (red star in Fig.~\ref{fig:time_domain}). 

After initializing and setting the external flux value at $\Phi_\mathrm{ext}^\mathrm{ref} + \delta\Phi_\mathrm{ext}^\mathrm{ref}$, the fourth step is to observe the decay of the flux states. We perform single-shot measurements by tracking the amplitude of the transmission value $|S_{21}|$ as a function of time at a read-out frequency of $f_\mathrm{rout} = 6.7321$ GHz. This value corresponds to the resonance frequency when the state is in the lower-lying valley. Thus, whenever the state decays, we detect a jump in the transmission value. Figure \ref{fig:time_domain} (b) shows examples of these jumps at the flux detuning value of $\delta\Phi_\mathrm{ext}^\mathrm{ref}=21.5$\,m$\Phi_0$ with the longest lifetime of over one hour. 

We perform the same time-domain measurements at increasingly larger flux values, which result in increasingly lower barrier heights. Depending on the barrier height, we observe relaxation times ranging from a few tens of seconds up to above an hour as the barrier height increases. Figure \ref{fig:histograms} presents histograms that show the time distribution of these jumps. 

\section{Conclusion}
\label{sec:conclusion}

In this work, we have demonstrated the nonlinear behavior of a WSi nanowire by incorporating it as a weak link element in an RF-SQUID with a three-dimensional cavity. The disordered superconductor WSi has been extensively used in superconducting nanowire single-photon detectors due to its extreme structural homogeneity and lack of grain boundaries~\cite{marsili_detecting_2013,oripov_superconducting_2023}. Additionally, WSi has been successfully incorporated in coherent quantum devices as a linear inductor~\cite{Larson2026}. By employing this material as a weak link element in an RF-SQUID and measuring its resonance condition, we have demonstrated its nonlinear behavior and modeled it as a Josephson junction with a sawtooth-like current-phase relation and a quantum phase slip element. We also measure the relaxation times of the excited metastable states trapped in local minima of the potential via single-shot readout. From several hundred single-shot readout measurements, we show that the relaxation times are controllable via tuning external flux. These results illustrate the potential use of nanobridge weak links of WSi as a new nonlinear element for superconducting circuits.

\section*{Acknowledgments}
We thank Leonid Glazman and Vladimir Manucharyan for the inspiring discussions. The research was sponsored by the Army Research Office under Grant Number W911NF-22-1-0050. The views and conclusions contained in this document are those of the authors and should not be interpreted as representing the official policies, either expressed or implied, of the Army Research Office or the U.S. Government. The U.S. Government is authorized to reproduce and distribute reprints for Government purposes notwithstanding any copyright notation herein. Certain commercial materials and equipment are identified in this paper to foster understanding. Such identification does not imply recommendation or endorsement by the National Institute of Standards and Technology, nor does it imply that the materials or equipment identified are necessarily the best available for the purpose.
S.G.J. acknowledges support from the National Science Foundation Graduate Research Fellowships Program under Grant No. DGE 2040434. T.F.Q.L acknowledges support in part by an appointment to the NRC Research
Associateship Program at the National Institute of Standards and Technology, administered by the Fellowships
Office of the National Academies of Sciences, Engineering, and Medicine.

\appendix

\section{Fabrication}
\label{app:fabrication}

The device fabrication is similar to the process in Ref.~\cite{PRXQuantum.5.020309}. We began by solvent cleaning a 3-inch silicon wafer capped with a thermally grown $\text{SiO}_{2}$ layer with two rounds of sonication in acetone and isopropanol (subsequently referred to as a solvent clean). The WSi film was then co-sputtered with tungsten and silicon in an inert (argon) atmosphere with a manual AJA Sputtering Deposition System.  The film used to make this RF-SQUID device was $\approx 3.5$ nm thick and had a chemical composition of W$_{0.85}$Si$_{0.15}$, achieved with a constant chamber pressure of 1.2 mTorr during deposition. To avoid oxidation, the film was capped with 2 nm of amorphous Si, sputter deposited \textit{in-situ} at a chamber pressure of 9 mTorr. 

The WSi wire was first patterned with optical lithography and subtractive plasma etching. We used a resist adhesion promoter (P20) beneath a $\approx$ 1 \textmu m layer of SPR 660 resist that we baked at 95 $^\circ$C for 1 minute on a hot plate before patterning it in an ASML 5500/100D wafer stepper. After exposure, the wafer was post-baked at 110 $^\circ$C for 1 minute and developed for 30 seconds in MF26A.  We performed an SF$_6$ plasma etch (20 sccm SF$_6$, 25 mTorr of pressure, 70 W, -140 V DC bias) in an ion and plasma equipment reactive ion etcher (IPE RIE). Pre-etch, the chamber was conditioned with a 10 minute O$_2$ clean and a 5 minute SF$_6$ pre-condition.  After etching, the wafer underwent another solvent clean to strip the remaining photoresist. 

Next, the aluminum capacitor pads in the circuit were fabricated.  We spun a 0.2 \textmu m thick layer of LOR 3A resist on top of a P20 adhesion layer, followed by baking for 5 minutes at 150 $^\circ$C. After baking, we spun a layer of SPR 660 resist with the same recipe as in the previous step (excluding an additional P20 layer) and exposed the new pattern in the ASML 5500/100D stepper.  Post exposure, we baked for 1 minute at 110 $^\circ$C and developed using MF26A before rinsing. After patterning, we deposited a 100 nm film using an Angstrom electron-gun evaporator with an \textit{in-situ} argon plasma RF clean prior to deposition.  {The RF clean was at a power of 30 W for 1 minute at a chamber pressure of 10 mT.} We then complete liftoff with PG remover overnight at room temperature before the wafer is finally rinsed with isopropyl alcohol (IPA).  

The aluminum loop for the meandering inductor and contact to the WSi nanowire was fabricated next.  We spun a 250 nm thick layer of PMMA and baked it at 180 C for 1 minute.  Then, the loop and its 100 nm gap to define the nanowire length were exposed using a JEOL e-beam lithographer. We develop in a methyl isobutyl ketone:isopropyl alcohol (MIBK:IPA) 1:3 solution for 1 minute, then RF clean to ensure galvanic contact with the previous aluminum layer and the WSi before depositing 100 nm of aluminum.  After deposition, we complete liftoff with PG remover similar to the previous step.

Finally, we defined the width of the WSi nanowire using e-beam lithography. We again spun a 120 nm thick layer of PMMA and baked it at 180 C for 1 minute. Then, we wrote the 100 x 40 nm WSi nanowire pattern in the JEOL electron beam lithography system, and then developed it in MIBK:IPA 1:3, similar to the previous step. The etch is completed in the Oxford IPE RIE with endpoint detection to avoid deep trenching in the silicon substrate.  After a final solvent clean, the wafer is diced and prepared for measurement. We note that after fabrication, the device was exposed to air for severarements took place.

\section{Circuit quantization}
\label{app:circuit_quantization}

\begin{figure}
    \centering
    \includegraphics[width = 8.6cm]{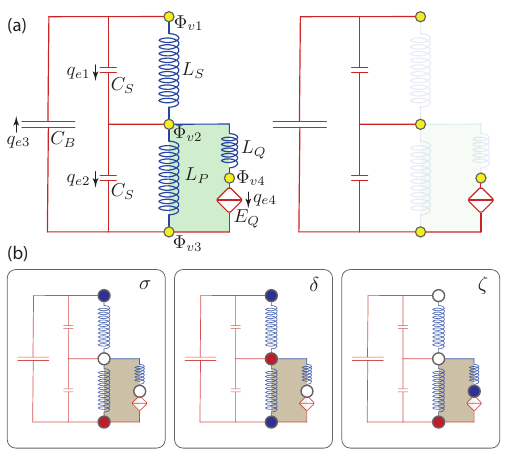}
    \caption{\label{fig:app_theory} (a) The capacitive subgraph of the circuit highlighted by red edges. (b) The three modes of the circuit. The colors of the node refer to the sign of how a node appears in the definition of the three modes.}
\end{figure}

Here, we detail how we arrive at the Hamiltonian of the circuit containing the quantum phase slip element [see Eq.~\ref{eq:qps_hamiltonian}], following the theoretical treatment discussed in Ref.~\cite{PRXQuantum.5.020309}. First, we identify the capacitive subgraph of the circuit [highlighted by red lines in Fig.~\ref{fig:app_theory}(a)], and write down the corresponding incidence matrix

\begin{equation}
\Omega_{ev} = 
\begin{pmatrix}
-1 & 1 & 0 & 0 \\
0 & -1 & 1 & 0 \\
1 & 0 & -1 & 0 \\
0 & 0 & 1 & -1
\end{pmatrix}.
\end{equation}
Here, the rows correspond to the branch (edge) charge variables $q_{ei}$ ($i=1,\dots4$), the columns to the flux node (vertex) variables $\Phi_{vj}$ ($j=1,\dots4$), and the value of an element refers to how a directed edge is connected to a vertex: +1 (-1) refers to an edge pointing into (out of) a vortex, and 0 indicates no connection. The total energy of the capacitive circuit elements is $E_\mathcal{C}(q_{ei})$, while the inductive energy is $E_\mathcal{I}(\Phi_{vj})$. With these definitions, the Lagrangian of the circuit reads
\begin{equation}
\begin{split}
    L &= \sum_{i,j} q_{ei}[\Omega_{ev}]_{ij}\dot\Phi_{vj} -\sum_iE_\mathcal{C}(q_{ei}) - \sum_jE_\mathcal{I}(\Phi_{vj})= \\
    &q_{e1}\left(\dot{\Phi}_{v2}-\dot{\Phi}_{v1}\right) + q_{e2}\left(\dot{\Phi}_{v3}-\dot{\Phi}_{v2}\right)+\\
    &q_{e3}\left(\dot{\Phi}_{v1}-\dot{\Phi}_{v3}\right)+q_{e4}\left(\dot{\Phi}_{v3}-\dot{\Phi}_{v4}\right) - \\&\sum_iE_\mathcal{C}(q_{ei}) - \sum_jE_\mathcal{I}(\Phi_{vj}).
\end{split}
\end{equation}

To remove the excess degrees of freedom, we notice that there is a left null vector of the incidence matrix, which corresponds to Kirchhoff's voltage law
\begin{equation}
\frac{q_{e1}}{C_S} + \frac{q_{e2}}{C_S} + \frac{q_{e3}}{C_B} = 0.
\end{equation}
After removing the excess variables, we introduce the following flux and charge variables [Fig.~\ref{fig:app_theory}(a)]
\begin{equation}
    \begin{split}
        \Sigma &= (\Phi_{v3} - \Phi_{v1}) / 2, \\
        \Delta &= (2\Phi_{v2}-\Phi_{v3} - \Phi_{v1}) / 2,\\
        Z &= (\Phi_{v3} - \Phi_{v4}), \\
        q_\Sigma &= q_{e1} + q_{e2} - 2 q_{e3}\\
        q_\Delta &= q_{e1} -q_{e2}, \\
        q_Z &= q_{e4}.\\
    \end{split}
\end{equation}

In this choice of variables, the Lagrangian reads
\begin{equation}
    L = q_\Sigma\cdot\dot{\Sigma} + q_\Delta\cdot\dot{\Delta}+q_Z\cdot\dot{Z} - E_\mathrm{tot}(q_\Sigma, q_\Delta, q_Z, \Sigma, \Delta, Z),
\end{equation}
where $E_\mathrm{tot}$ is the sum of the capacitive and inductive energies.

Finally, the Hamiltonian of the circuit takes the form
\begin{equation}
\begin{split}
    H &= \frac{q_\Delta^2}{4C_S} + \frac{q_\Sigma^2}{4(2C+C_S)}-E_Q\cos\left(2\pi \frac{q_Z}{2e}\right) \\
    +&\frac{1}{2L_S}\left(\Sigma+\Delta\right)^2+\frac{1}{2L_P}\left(\Sigma-\Delta-\Phi_\mathrm{ext}\right)^2 \\
    &+\frac{1}{2L_Q}\left(\Sigma-\Delta-Z\right)^2.
\end{split}
\end{equation}
We arrive at the Hamiltonian in Eq.~\ref{eq:qps_hamiltonian} after introducing the phase variables $\sigma=2\pi\Sigma/\Phi_0$, $\delta=2\pi\Delta/\Phi_0$, $\zeta=2\pi Z/\Phi_0$, and the conjugate Cooper pair number variables $n_\sigma=q_\Sigma/2e$, $n_\delta=q_\Delta/2e$, $n_\zeta=q_Z/2e$, and promoting the classical variables into quantum operators.

We solve for the eigenenergies of this Hamiltonian using an analytical approach. First, we remove the light, high-energy $\delta$-mode using the Born-Oppenheimer approximation~\cite{Paolo_2019}. We note that the potential energy at the various $\zeta$ values resembles a two-dimensional harmonic potential well [Fig.~\ref{fig:theory}(d)]. By using the Born-Oppenheimer approximation, we approximate these two-dimensional wells with a harmonic well in one dimension. For example, at $\zeta=0$ and $\Phi_\mathrm{ext}=0$,  we introduce the following Hamiltonian, where $\sigma_0$ is a parameter 
\begin{equation}
\begin{split}
\hat{H}_\delta &=  4E_{C\delta}\hat{n}_\delta^2 + 
\frac{1}{2}E_{LS}\cdot\left(\sigma_0+\hat{\delta}\right)^2 + \\& \frac{1}{2}E_{LP}\cdot\left(\sigma_0-\hat{\delta}\right)^2 +  \frac{1}{2}E_{LQ}\cdot\left(\sigma_0-\hat{\delta}\right)^2.
\end{split}
\end{equation}
Then, to obtain the one-dimensional approximation of the problem, we combine the ground state energy $E_0(\sigma_0)$ of $\hat{H}_\delta$ as an effective potential with the kinetic energy of the heavy mode
\begin{equation}
\begin{split}
\hat{H}_\sigma &=  4E_{C\sigma}\hat{n}_\sigma^2 + E_0(\hat\sigma).
\end{split}  
\end{equation}

\begin{figure}[h]
    \centering
    \includegraphics[width = 8.6cm]{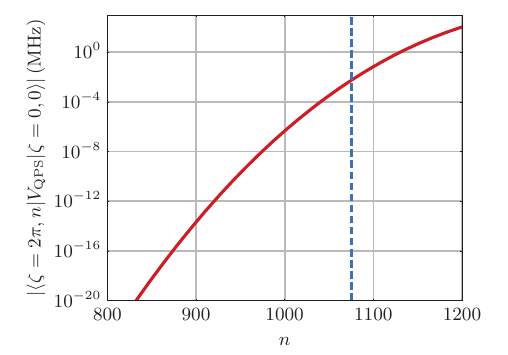}
    \caption{\label{fig:matrixelements}  {The coupling matrix elements between the neighboring valleys as a function of the energy index.}}
\end{figure}

This way, we obtain a series of one-dimensional harmonic potentials (with the same curvature) at integer multiples of $2\pi$ of $\zeta$ values that can host harmonic oscillator states $|\zeta,n\rangle$, where $n$ refers to the $n$th excited state of the oscillator with energy $E_n$. The harmonic oscillators are coupled through the quantum phase slip term $\hat{V}_\mathrm{QPS}=- E_\mathrm{WSi}^\mathrm{QPS}\cos\left(2\pi\hat{n}_\zeta\right)$. We use second-order perturbation theory, where the phase slip term $\hat{V}_\mathrm{QPS}$ is treated perturbatively, to obtain the energies of the full system. In this case, the change of the energy of the $|\zeta=0,n\rangle$ state due to one of its neighbors is
\begin{equation}
    \Delta E_{0,n}=\sum_{m\neq n}\frac{\left|\langle\zeta=2\pi,m|V_\mathrm{QPS}|\zeta=0,n\rangle\right|^2}{E_n-E_m}.
\end{equation}
We use the analytical properties of harmonic oscillator wavefunctions (using equation 7.377 in~\cite{Gradshtein_Ryzhik_2007}) to evaluate the matrix elements and obtain the curve represented in Fig.~\ref{fig:theory}(f).  {The magnitude of these coupling matrix elements as a function of $n$ is shown in Fig.~\ref{fig:matrixelements}. The blue dashed curve indicates the energy level index, where the state in the lower valley is resonant with the lowest energy state in the upper valley at the critical flux value.}

\section{Higher energy levels}
\label{app:full_spectrum}

\begin{figure}[t]
    \centering
    \includegraphics[width = 8.6cm]{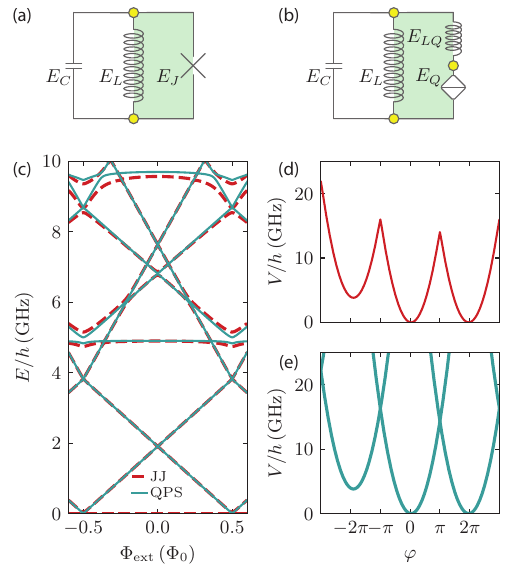}
    \caption{\label{fig:higher_levels}  {(a), (b) Circuit schematics of the JJ- and QPS-based models. (c) Numerical calculation of the energy spectrum for the two cases. (d), (e) Potential landscape of the two circuits. The parameters used are the following: $E_C/h=1\,$GHz, $E_L/h=0.1\,$GHz, $E_J/h=E_{LQ}/h=3\,$GHz, $E_Q/h=5\,$GHz, and the skewness is 0.99.}}
\end{figure}

 {Here, we provide more information on how one could potentially distinguish between the JJ- and QPS-based models using higher energy levels. In Fig.~\ref{fig:higher_levels}(a) and (b), we show the simplified version of the studied circuits, and calculate the full energy spectrum of such devices. As Fig.~\ref{fig:higher_levels}(c) illustrates, the low energy part of the circuits is indistinguishable, while, as the energy of the levels is increased, the difference becomes more apparent. This can be understood by investigating the different potential energy landscapes of the circuits [Fig.~\ref{fig:higher_levels}(d), (e)]. For the JJ-based circuit, the potential energy is one-dimensional, while for the QPS-based circuit, it is two-dimensional with multiple shifted parabolas. At low energies, the potential landscapes exhibit similarities, but at higher energies, the differences become more pronounced.}

\section{Finite-element modeling of the circuit}
\label{app:finite-element}

We use ANSYS Maxwell and HFSS to extract the circuit parameters and bare resonance frequencies of the coupled RF-SQUID and cavity system. The orientation of the chip inside the cavity is such that it lies in the $xy$-plane with the capacitor pads parallel to the longest side of the cavity ($y$ axis), and the shortest side of the cavity is parallel to the $x$ axis. In this case, the lowest-lying modes of the bare cavity are the TE$_{101}$, TE$_{201}$, and TE$_{301}$ modes, out of which TE$_{201}$ has a node in the middle of the cavity, where the chip is located, and we ignore its effects. For abbreviation, we call the TE$_{101}$ and TE$_{301}$ modes as mode $a$ and $b$. Table \ref{tab:bare_system} summarizes the simulated (on a model adjusted by the machining imperfections) and measured resonance frequencies (at room temperature) of the bare cavity.

\begin{table}[h!]
\begin{center}
\caption{Simulated and measured frequencies of the modes of the bare cavity.}
\label{tab:bare_system}
\begin{tabular}{ |c|c|c|c| } 
 \hline
  & $f_{\mathrm{TE}_{101}}$ (GHz)& $f_{\mathrm{TE}_{201}}$ (GHz)& $f_{\mathrm{TE}_{301}}$ (GHz)\\ 
\hline
 sim. & 9.07 & 11.83 & 15.37 \\ 
 meas. & 9.08 & 11.79 & 15.30 \\ 
 \hline
\end{tabular}
\end{center}
\end{table}

To simulate the coupled system, we substitute the nanowire with a small geometric inductance of $L^\prime=134$\,pH, and place the RF-SQUID, including the silicon substrate, into the model cavity. We find the relevant resonance frequencies of the coupled system ($f_0$, $f_1$, $f_2$) that are presented in Table \ref{tab:coupled_system}. Note that the lowest-lying frequency closely matches the experimental data (6.731 GHz), while the other two frequencies are out of the band of the amplifying chain of the measurement. In addition, we note that the resonance frequency of the TE$_{201}$ mode only changes slightly from 11.83 GHz to 11.50 GHz, when the chip is inserted in the cavity, because that mode has a node around the location of the chip, justifying that we can ignore its effects.

\begin{table}[h!]
\begin{center}
\caption{Simulated and measured frequencies of the coupled system.}
\label{tab:coupled_system}
\begin{tabular}{ |c|c|c|c| } 
 \hline
  & $f_0$ (GHz) & $f_1$ (GHz) & $f_2$ (GHz) \\ 
\hline
 sim. & 6.68 & 9.67 & 13.86 \\ 
 meas. & 6.73 & - & - \\ 
 \hline
\end{tabular}
\end{center}
\end{table}

We model the full system as an RF-SQUID resonator coupled to modes $a$ and $b$. While we know the resonance frequency of the modes in the empty cavity, placing the RF-SQUID inside the cavity modifies the bare resonance frequencies of the modes due to additional capacitive pads in the cavity. We use the simulation to find these renormalized cavity frequencies ($f_{a}$ and $f_{b}$). The loaded resonance frequency of the RF-SQUID $f_{q}$ can be obtained by simulating the capacitance (including the capacitances of the device to the wall of the cavity) and the geometrical inductance of the device, from which we find $f_{q}=$ 10.08 GHz in the lumped element approximation. The full coupled system is described by the following Hamiltonian
\begin{equation}
\begin{split}
H =hf_q \hat{c}^\dagger \hat{c} + hf_a \hat{a}^\dagger \hat{a} + hf_b \hat{b}^\dagger \hat{b} \\
+ \hbar g_a (\hat{c}^\dagger \hat{a} + \hat{a}^\dagger \hat{c}) + \hbar g_b (\hat{b}^\dagger \hat{c} + \hat{c}^\dagger \hat{b}),
\end{split}
\end{equation}
where $\hat{a}$, $\hat{b}$, $\hat{c}$ are the annihilation operators of the modes and the RF-SQUID, and the coupling between the modes and the RF-SQUID are $g_a$, and $g_b=\sqrt{\frac{\omega_b}{\omega_a}}g_a$. 

By fitting the three resonance frequencies from the simulation with the eigenenergies of the full system Hamiltonian, we obtain the values summarized in Table \ref{tab:renormalized_system}. In the numerical calculations, we use the simulated values of $f_a$ and $f_b$, and vary the coupling to match the experimental data.

\begin{table}[H]
\begin{center}
\caption{Renormalized frequencies of the cavity and coupling values.}
\label{tab:renormalized_system}
\begin{tabular}{ |c|c|c|c| } 
 \hline
  $f_a$ (GHz) & $f_b$ (GHz) & $g_a/2\pi$ (GHz) & $g_b/2\pi$ (GHz) \\ 
\hline
  8.07 & 12.02 & 1.89 & 2.26 \\ 
 \hline
\end{tabular}
\end{center}
\end{table}

\section{Flux calibration}
\label{app:flux calibration}

\begin{figure}[h]
    \centering
    \includegraphics[width = 8.6cm]{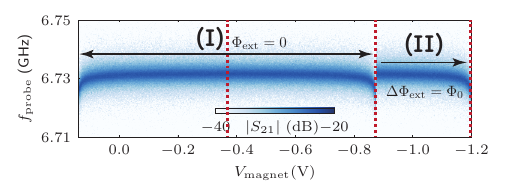}
    \caption{\label{fig:flux}  {Schematics of the external flux calibration procedure based on the measured data. The two steps are highlighted, and the dashed lines show the zero magnetic field reference point, and the jumping values.}}
\end{figure}

 {Figure \ref{fig:flux} shows the two steps we use to calibrate the relation between the applied voltage supplied to the external magnet and the external flux threading the loop of the device. (I) First, we find the zero magnetic field point by increasing and decreasing the magnetic field in a way that we approach the first flux jumps at positive and negative flux values (without inducing a flux jump). We define the symmetry point in the middle as the reference zero flux value. (II) Then we increase the magnetic field, and measure the voltage period of flux jumps. Given that we have observed the same period of jumps over several different runs, and the fact that in dissipative systems, the jumps are most likely to occur between the neighboring valleys, we conclude that this voltage period corresponds to a change in magnetic field with a single flux quantum. This allows us to find the scaling factor between the applied voltage and the flux.}

\section{Fridge Measurement Setup}
\label{app:fridge-measurement}

\begin{figure}[H]
    \centering
    \includegraphics[width=\textwidth/2]{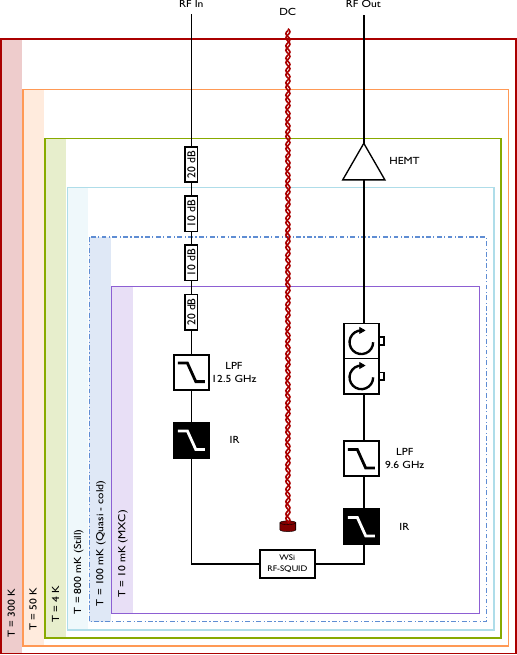}
    \caption{\label{fig:fridgewiring} Wiring diagram of the dilution fridge measurement setup.}
\end{figure}

A diagram of our fridge wiring setup is shown in Fig.~\ref{fig:fridgewiring}. The low-pass filters on the input and output lines have cuttoff frequencies of 12.5 and 9.6 GHz, respectively. There are two Eccosorb filters for blocking infrared signals directly before and after the RF-SQUID. For all measurements, the RF-SQUID and 3D cavity were mounted to a copper cold finger on the mixing chamber flange and housed in a light-tight aluminum shield nested inside a $\mu$-metal shield. The flux for measurements was applied using an electromagnet that was mounted to the RF-SQUID housing.

The room temperature electronics consisted of two general configurations: one for continuous-wave measurements and one for time-domain measurements. The continuous-wave measurements were taken using a vector network analyzer.  The time domain measurements were completed using the Quantum Machines OPX system and external local oscillators. 

\end{document}